\documentclass[12pt]{article}
\usepackage[latin1]{inputenc}
\usepackage[reqno]{amsmath}
\usepackage{amssymb,latexsym}
\usepackage{multicol}
\usepackage{caption}
\usepackage{float}
\usepackage{graphicx}
\usepackage{color} 
\newcommand{\cor}{\color{black}}

\usepackage[sort, numbers, super, compress]{natbib}
\usepackage{indentfirst}
\usepackage{epstopdf}
\title{\textbf{Forward Flux Sampling {\cor Calculation} of Homogeneous Nucleation Rates from Aqueous NaCl Solutions}}
\author{Hao Jiang$^{\dag}$, Amir Haji-Akbari$^{\ddag}$, Pablo G. Debenedetti$^{\dag}$, \\ and Athanassios Z. Panagiotopoulos$^{\dag}$\footnote{Corresponding author. E-mail: azp@princeton.edu.} \\
\footnotesize\it$^{\dag}$ Department of Chemical and Biological Engineering, Princeton University, \\ 
\footnotesize\it Princeton, New Jersey 08544, United States \\
\footnotesize\it$^{\ddag}$ Department of Chemical \& Environmental Engineering, Yale University, \\ \footnotesize\it New Haven, CT, 06520, United States \\}

\date{}
\begin{document}
\maketitle 

\centerline{\textbf{Abstract}}
We used molecular dynamics simulations and {\cor the} path sampling technique known as forward flux sampling to study homogeneous nucleation of NaCl crystals from supersaturated aqueous solutions {\cor at 298 K and 1 bar}. Nucleation rates were obtained for a range of salt concentrations for the Joung-Cheatham NaCl force field combined with the SPC/E water model. The calculated nucleation rates are significantly lower than available experimental measurements. The estimates for the nucleation rates in this work do not rely on classical nucleation theory, but the pathways observed in the simulations suggest that the nucleation process is better described by classical nucleation theory than an alternative {\cor interpretation based on} Ostwald's step rule, in contrast to some prior simulations of related models. In addition to the size of NaCl nucleus, we find that the crystallinity of a nascent cluster plays an important role in the nucleation process. Nuclei with high crystallinity were found to have higher growth probability and longer {\cor lifetimes}, possibly because they are less exposed to hydration water. 

\section{Introduction}
How solute molecules precipitate out of solution to nucleate into solid crystals is a fundamental scientific question with important ramifications in many disciplines. For instance, controlling the nucleation process is extremely important in pharmaceutical and petroleum industries;\cite{Chen, Gupta} while the precipitation and crystallization of electrolytes from supersaturated aqueous solutions is relevant to atmospheric sciences, biology and geochemistry.\cite{Driessche} Despite its importance, there is a significant gap in our understanding of electrolyte precipitation as the highly non-equilibrium nature of the nucleation process, which involves the formation of short-lived nanometer-scale intermediates, makes probing the microscopic mechanism of nucleation extremely difficult.\cite{Sosso} In other words, the existing experimental techniques lack the spatiotemporal resolution necessary for detecting critical nuclei.\cite{Debenedetti} While molecular simulations do not suffer from such lack of resolution, nucleation, as a rare event, usually occurs at timescales considerably longer than microseconds, {\cor for the typical system size ($N$ $\sim$  10$^3$) used in simulations. This fact, combined with the lack of thermal averaging provided by sampling single trajectory, places the direct calculation of nucleation rates beyond the reach of conventional molecular dynamics simulations. }

In this work, we focus on the nucleation of the NaCl crystal from supersaturated aqueous solutions. NaCl is the most abundant salt in seawater. It has also been extensively studied computationally, and there are several available force fields {\cor that yield accurate} solution and crystal chemical potentials.\cite{Nezbeda} These chemical potentials in turn specify the driving forces for the homogeneous nucleation.\cite{Benavides} There have been a few simulation studies of nucleation of NaCl crystals in supersaturated aqueous solutions. Mucha and Jungwirh studied evaporation-induced nucleation of NaCl using realistic interaction potentials\cite{Mucha} and molecular dynamics (MD) simulations. In another early simulation study, Zahn \cite{Zahn} used the transition path sampling method to analyze the initial stages of nucleation; it was concluded that the formation of non-hydrated sodium is important in the nucleation process.  Alejandre and Hansen found  clustering of NaCl ions and the crystallization mechanism to be highly sensitive to small changes in force field parameters.\cite{Alejandre} While these prior simulation studies revealed important features of the crystallization of NaCl in supersaturated solutions, they were conducted in closed ensembles (NPT or NVT) with simulation boxes consisting only a few hundred particles (water molecules, Na and Cl ions). It is therefore likely that in all such studies, the degree of supersaturation of the simulated aqueous solutions decreased during the formation of the crystalline phase. Since solution supersaturation is the driving force for nucleation, a depleted solution may not necessarily reflect the actual nucleation mechanism. 

In order to eliminate the effect of solution depletion, open-ensemble (e.g. Grand-Canonical\cite{Eslami} or osmotic ensemble\cite{Lisal}) simulations that constrained system chemical potentials could be used, however the effect of inserting or removing solute/solvent particles on the nucleation process requires further analysis. Alternatively, closed-ensemble NPT or NVT simulations with a large system size can be used to minimize such depletion effects. Chakraborty and Patey performed such large-scale MD simulations of nucleation in aqueous NaCl solutions using the SPC/E water\cite{Berendsen} and OPLS NaCl\cite{Chandrasekhar, Aqvist} force fields. \cite{Patey1, Patey2} They found that the nucleation of NaCl follows a two-step mechanism, which involves the formation of ionic aggregates followed by rearrangement of clusters into structured crystals. Their findings are in contrast to the one-step mechanism predicted by classical nucleation theory (CNT). The system simulated by Chakraborty and Patey {\cor had} a salt concentration of 4 mol/kg. We, however, found the solubility of the OPLS NaCl in SPC/E water to be less than 0.02 mol/kg {\cor at 298 K and 1 bar} (details given in Supplementary Material). Therefore, the supersaturation ratio, defined as solution concentration divided by the equilibrium solubility, is more than 200 in the Chakraborty and Patey's simulations, a value significantly higher than the experimentally measured metastability limit,\cite{Desarnaud} raising questions about the relevance of the two-step nucleation mechanism. More recently, Lanaro and Patey used the SPC/E water and Joung-Cheatham (JC) NaCl\cite{Joung} force fields to study the effects of nucleus size and crystallinity on the NaCl nucleation process, and concluded that the crystallinity has crucial {\cor impact} on the stability and probability of achieving nucleation for an ionic cluster.\cite{Lanaro}

In recent years, there has been an increased {\cor interest} in using enhanced sampling techniques to study crystal nucleation. One such technique is metadynamics, which was recently applied to study NaCl nucleation in solution by Giberti {\it et al.}\cite{Giberti}. They showed that the NaCl nucleation pathway may involve wurtzite (tetrahedral) structure, consistent with the Ostwald step rule.\cite{Santen} However, it was later found that the Gromos force field employed in the metadynamics study exaggerates the stability of the wurtzite phase.\cite{Zimmermann} Thus, the question of homogeneous nucleation mechanism for NaCl from aqueous solution remains open. 

Despite numerous computational studies of NaCl nucleation, most such studies have been performed at relatively high NaCl concentration, so that spontaneous nucleation may be observed within the time scale of a MD simulation, i.e. a few hundreds nanoseconds. In particular, the nucleation rates have not generally been obtained for a sufficiently wide range of salt concentrations except in the work by Zimmermann {\it et al.},\cite{Zimmermann} who used the CNT-based seeding technique to estimate the critical nucleus size, effective interfacial tensions, ion-attachment kinetics, and subsequently the nucleation rates. They found that the SPC/E and JC force fields overestimate nucleation rates in comparison with available experimental data. It has, however, been recently demonstrated that the NaCl solubility of 5.1 mol/kg\cite{Aragones} utilized by Zimmermann {\it et al.} in estimating the thermodynamic driving force for nucleation is an overestimation, with the correct value being 3.7 mol/kg.\cite{Espinosa} This inevitably leads to an underestimation of an nucleation barrier and an overestimation of rate. Also, it is not entirely clear if the nucleation of NaCl in supersaturated aqueous solution can be described by CNT, especially in light of prior simulation studies suggesting the possibility of a two-step nucleation mechanism. 

In the present study, we used the forward flux sampling (FFS) method\cite{Allen1, Allen2} to estimate the homogeneous nucleation rates of NaCl in aqueous solutions over a range of supersaturations. The FFS method has been successfully applied to study a series of rare events that are relevant to crystallization, including homogeneous and heterogeneous nucleation of ice,\cite{Akbari1, Akbari2, Bi1} and other tetrahedral liquids,\cite{Galli,Gianetti} nucleation of NaCl crystals from melt,\cite{Valeriani} crystallization of methane hydrate,\cite{Bi2} as well as nucleation of model systems.\cite{Filion,Speck} With the FFS framework, nucleation rates can be determined directly, without any need to invoke classical nucleation theory. {\cor Furthermore, because FFS involves the generation of a large number of trajectories, it is possible to gather relevant information on mechanisms.\cite{Akbari1}}

The paper is structured as follows: in section II we briefly discuss the utilized molecular force fields and provide details of our FFS simulations. In section III, we report the homogeneous nucleation rates obtained from FFS, and discuss mechanistic details such as the nucleation pathways, {\cor as well as the} dipole moments, dimensionality, and asphericity of the crystalline nuclei. We also discuss the effects of crystallinity and hydrated water content of a nucleus on its stability and growth probability. The free energy profile of NaCl {\cor clusters} is also investigated {\cor as a function of their size}. Finally, conclusions are summarized in section IV.

\section{Models and Simulation Details}
We studied the nucleation of NaCl in solutions with salt concentrations ranging from 8.0 to 16.6 mol/kg {\cor at 298 K and 1 bar}. The SPC/E water\cite{Berendsen} and JC NaCl\cite{Joung} force fields were used in all MD simulations. The solubility  of the JC model of NaCl in the SPC/E water is 3.7 mol/kg {\cor at 298 K and 1 bar}, a value that has been confirmed by both chemical potential calculations\cite{Mester} and direct coexistence simulations.\cite{Espinosa}  In addition, the SPC/E model in conjunction with the JC force field, provides reasonable prediction for several thermophysical properties of aqueous NaCl solutions,\cite{Orozco} and has been previously used for simulating NaCl crystal nucleation.\cite{Zimmermann, Lanaro} Further details about the force field parameters can be found in the Supplementary Material. To quantify the driving force of the nucleation, we extend the calculation of NaCl chemical potential in solution to supersaturated conditions up to 16 mol/kg following the method proposed by Mester and Panagiotopoulos.\cite{Mester} The calculated chemical potential {\cor of NaCl in solution} as a function of salt concentration is shown in the Supplementary Material. Unless stated otherwise, the simulated systems consisted of 1000 Na$^+$ and 1000 Cl$^-$ ions, along with 3600 to 5550 water molecules for the higher {\cor concentrations}, and 2000 ion pairs with 13900 water molecules for the lowest (8 mol/kg). We found critical nuclei to be smaller than 120 ions, as discussed in the next section. This is only a small fraction of the total number of ions in the system, thus ensuring that the degree of supersaturation does not change significantly during the nucleation process. 

Forward flux sampling (FFS) was applied to obtain homogeneous nucleation rates. The FFS method samples the nucleation process at a series of milestones defined by an order parameter. We used the number of ions (Na$^+$ and Cl$^-$) in the largest crystalline nucleus as the order parameter $\lambda$. The local q4 {\cor bond orientational order} parameter\cite{Steinhardt} was used to distinguish between the ions in solution and those in the crystalline domains. In particular, two ions were considered as neighbors if their distance was less than 0.4 nm, almost coinciding the first minimum of the Na-Cl radial distribution function in solution, and an ion ($i$) was labelled as crystalline if it had at least 4 neighbors and its neighbor-average q4 order parameter ($=1/N_B \sum_{j=1}^{N_B} \sum_{m=-4}^{4} q_{4m}^*(i)\*q_{4m}(j)$) was larger than 0.35. Two crystalline ions that were within 0.35 nm of {\cor each other} (about the first minimum of the Na-Cl radial distribution function in the crystal) were considered to be part of the same crystalline nucleus. A similar ion-labelling strategy was used in a recent study of direct-coexistence MD simulations for calculation of NaCl solubility.\cite{Espinosa} It is worth mentioning that nucleation rates calculated from the FFS method are not very sensitive to the particular choice of an order parameter.\cite{Allen1} The first milestone was placed in the solution basin ($\lambda_{basin}=2$), and we subsequently defined a set of $N$ non-overlapping milestones with $\lambda_{N} > \lambda_{N-1} > ... > \lambda_{1} > \lambda_{0} = \lambda_{basin}$ -- with the corresponding $\lambda$ values listed in Supplementary Material. 500-600 independent MD simulations (total of 0.1, 0.3, 4.7 and 2.7 $\mu s$ of simulation time at 15.4, 12.0, 10.0 and 8.0 mol/kg, respectively) were conducted in the solution basin to estimate the initial flux ($\Phi$) {\cor as $1.0/(Vt_{1})$,  where $V$ is the system volume and $t_{1}$ is the averaged time required for a trajectory originated from solution basin to reach $\lambda_{1}$}.

Simulations were initiated from different configurations in order to ensure a sufficient sampling of possible nucleation pathways and collection of statistically independent configurations. Simulations in the solution basin were terminated if the trajectories reached the {\cor second} milestone ($\lambda_{1}$), and the averaged time required for a system to reach $\lambda_{1}$ was used to estimate the initial flux {\cor as described above}. After simulations in the solution basin, we initiated a large number of trial MD trajectories from the configurations collected at {\cor $\lambda_{j}$} by randomizing their momenta according to the Maxwell-Boltzmann distribution, and terminated those trajectories if they reached the next milestone ($\lambda_{j+1}$) or returned back to the solution basin. The transition probability $P(\lambda_{j+1}|\lambda_{j})$ is calculated as the ratio of the number of trial MD runs that successfully reach milestone {\cor $j$+1} and the total number of trial runs {\cor launched from milestone $j$}. We repeated this procedure until at a certain milestone ($\lambda_{N}$) the transition probability $P(\lambda_{N+1}|\lambda_{N})$ was 1.0, i.e. a crystal nucleus with $\lambda_{N}$ ions always grows. At each milestone, we ensured 150-300 successful crossings (except for the last two milestones where the transition probability was close to 1), in order to estimate accurately the transition probability. The nucleation rate ($J$) is the product of the initial flux and cumulative transition probability ($J = \Phi \times \sum_{i=0}^{N}P(\lambda_{i+1}|\lambda_{i})$). Values of the transition probabilities obtained are listed in the Supplementary Material.  

During the MD simulations, we performed an on-the-fly order parameter analysis every 2 ps using the open-source free energy sampling package PLUMED,\cite{Plumed} and all MD simulations were conducted in using GROMACS (version 5.1.4).\cite{gromacs} The timestep was 2 fs, and the cut-off distances for the van der Waals and real-space electrostatic interactions were both set to 0.9 nm. The particle-mesh Ewald summation method was used to handle long-range electrostatics with {\cor the Fourier spacing parameter set to 0.12 nm}.\cite{Darden} Temperature and pressure were controlled with the Nose-Hoover thermostat\cite{Nose} {\cor (time constant 2 ps)} and Parrinello-Rahman barostat\cite{Parrinello} {\cor(time constant 4 ps)} , respectively. The internal degrees of freedom of SPC/E water molecules were constrained using the LINCS algorithm.\cite{lincs}

\section{Results and Discussion}
The homogeneous nucleation rates for NaCl in supersaturated aqueous solutions were obtained using the FFS method at $m=$ 8.0, 10.0, 12.0 and 15.4 mol/kg {\cor at 298 K and 1 bar}, corresponding to a supersaturations ($S = m/m_{eq}$) of 2.1, 2.7, 3.2 and 4.1, respectively. Properties of the crystalline nuclei as well as {\cor cluster} free energy profiles were analyzed based on the configurations collected at FFS milestones.
 
\subsection{Critical Nucleus and Nucleation Rates}

\begin{figure}[H]
\centering
\includegraphics[width=10cm]{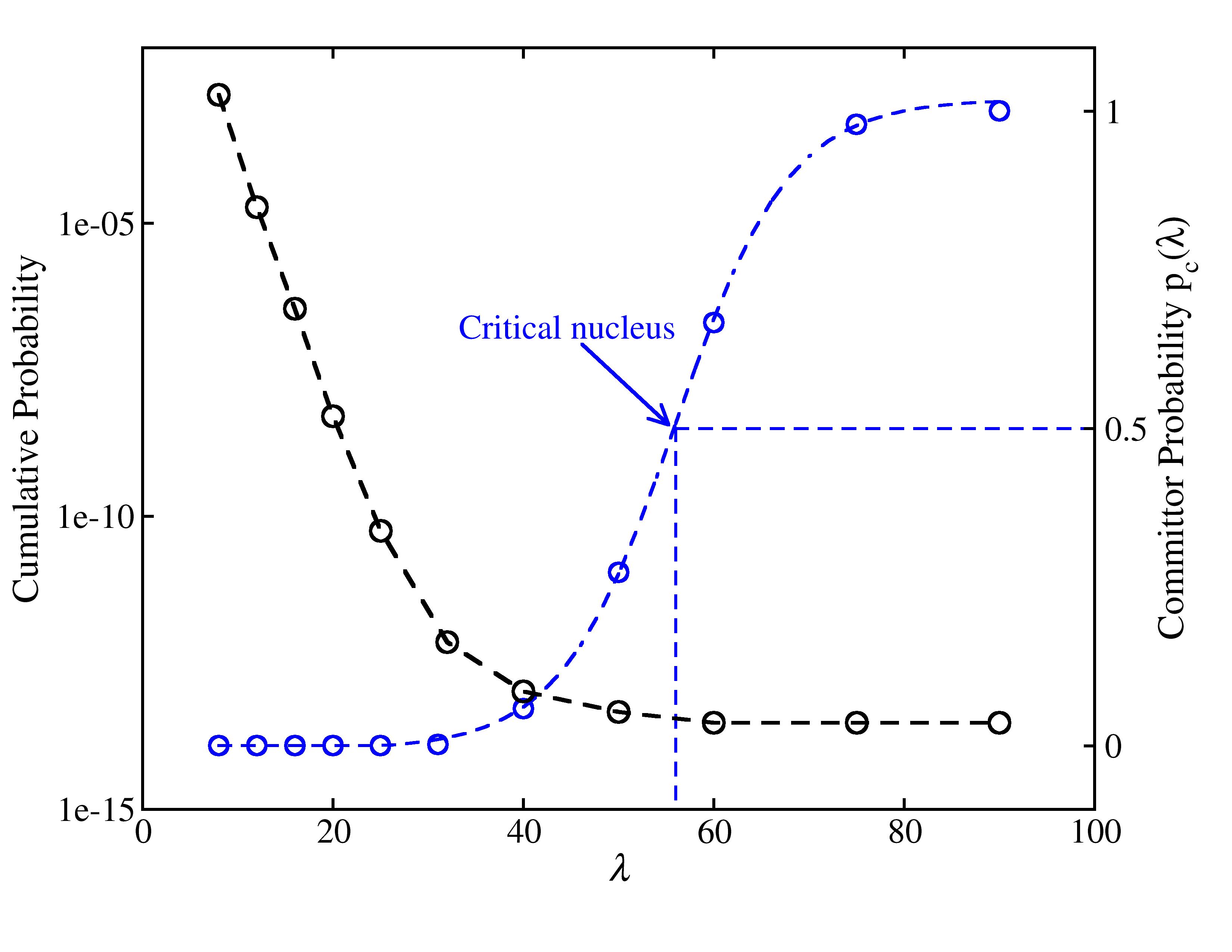}
\caption{Cumulative probability (black) and the committor probability $p_c(\lambda)$ (blue) as a function of order parameter ($\lambda$) at $S$ = 2.7 {\cor and at 298 K and 1 bar}. Dashed lines are guides to eyes. }
\label{FFS}
\end{figure}

Fig. \ref{FFS} shows the cumulative probability of the FFS simulations at supersaturation $S=m/m_{eq}=2.7$, corresponding to $m$=10 mol/kg for our model. The cumulative probability converged after the order parameter reached around 65, i.e. a crystalline nucleus with more than 65 ions always grows (has a transition probability of one), as can be seen in Figure \ref{FFS}. The calculated initial flux ($\Phi$) was $3.8\times10^{32}$ m$^{-3} $ s$^{-1}$, therefore the nucleation rate ($J$) was $1.1\times10^{19}$ m$^{-3} $ s$^{-1}$. The critical nucleus, which has a 50\% probability to grow to the crystalline domain (or return back to the solution phase), may be estimated from the commitor probability {\cor $p_c(\lambda_k)$} (= $\Pi_{j=k}^{N-1} P(\lambda_{j+1}|\lambda_{j})$). As shown in Fig. \ref{FFS}, the critical nucleus size, {\cor corresponding to $p_c$ = 0.5}, is around 56.

\begin{figure}[H]
\centering
\includegraphics[width=10cm]{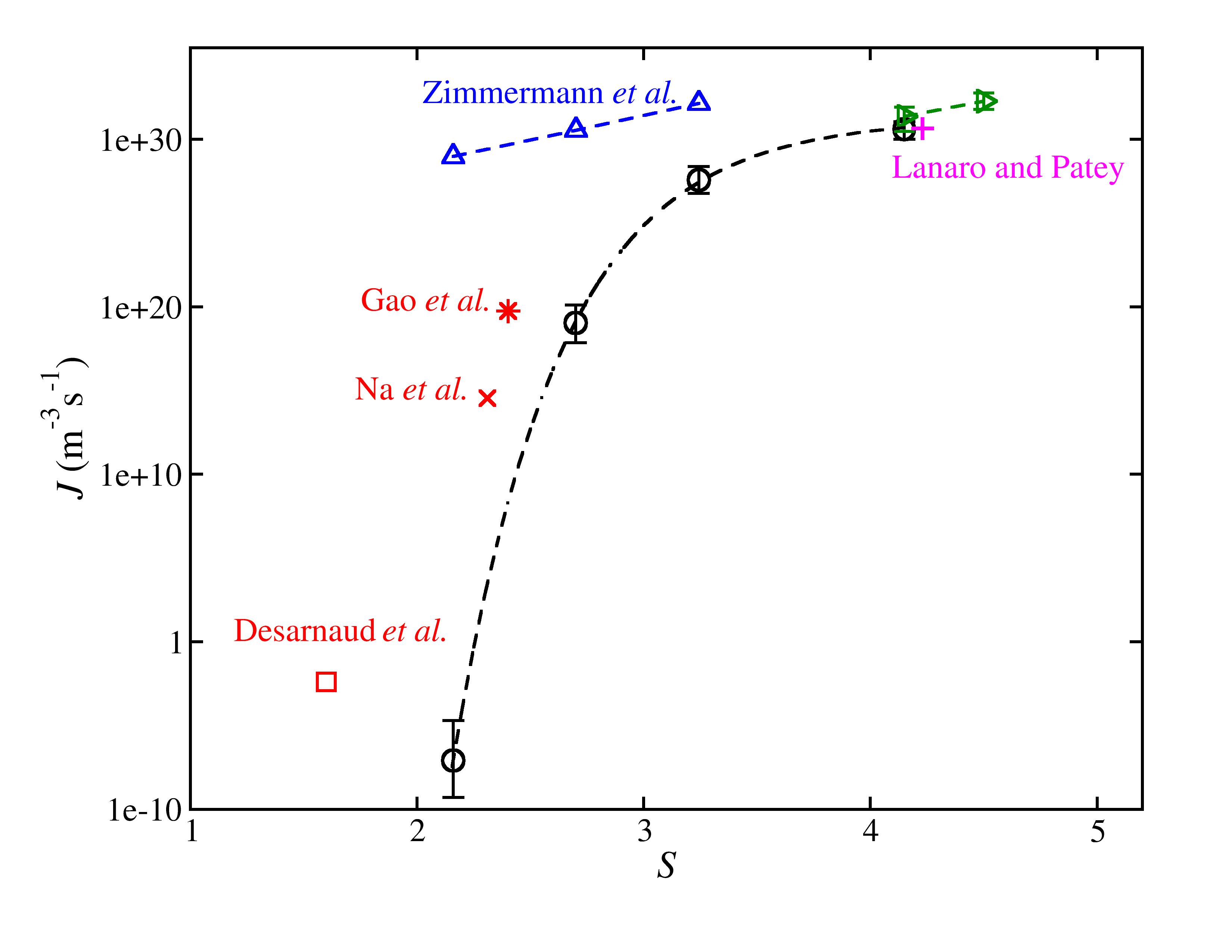}
\caption{Homogeneous nucleation rates ($J$) of NaCl in aqueous solutions at different supersaturations, $S = m/m_{eq}$. The {\cor open} black circles correspond to rates computed {\cor in this work}, while {\cor green triangles right} are the rates estimated from induction times of spontaneous nucleation in MD simulations. Red symbols are experimental estimation of nucleation rates from Na {\it et al.}\cite{Na}, Gao {\it et al.}\cite{Gao} and Desarnaud {\it et al.}\cite{Desarnaud} Magenta ``+'' is the nucleation rate estimated from induction time reported by Lanaro and Patey\cite{Lanaro}. Blue triangles are nucleation rates estimated by Zimmermann {\it et al.} \cite{Zimmermann} using the classical nucleation theory based seeding method. For spontaneous nucleation, the simulated systems had 1800 ions pairs with 6500 (at $S$=4.1, $m$=15.4) or 6000 (at $S$=4.5, $m$=16.6) water molecules. The uncertainties of nucleation rates from FFS simulations were estimated using the approach in Ref. \citenum{error}. }
\label{rate}
\end{figure}

In Fig. \ref{rate}, we report the nucleation rates ($J$) from our FFS simulations for supersaturations of 2.1, 2.7, 3.2 and 4.1. From the commitor probability, the critical nucleus sizes are 120, 56, 38, and 32, respectively, at these concentrations. At $S$ around 4.1, it was possible to observe spontaneous nucleation of NaCl within a few hundred nano-seconds of simulation time without applying the FFS method, and we obtained the nucleation rates at these high supersaturations (4.1 and 4.5) from the induction times averaged over 10 independent MD simulations that have undergone nucleation. As shown in the Figure S3 of the Supplementary Material, the metastability limit, which is defined as the salt concentration where the derivative of solution chemical potential with respect to salt concentration equals to 0, is around $S$ = 3.7 ($m$ = 14 mol/kg). Such finding is consistent with our observation that spontaneous nucleation happens at $S$ = 4.1 and 4.5, which are above the metastability limit. {\cor It should be noted, however, that a change in mechanism should occur for phase separation beyond a limit of stability (change form nucleation to spinodal decomposition), hence these observations of spontaneous ``nucleation'' should be interpreted with caution.} Note that at $S$ of 4.1, the discrepancy between the nucleation rate computed from FFS and the rate estimated from conventional MD is smaller than the statistical uncertainty of each calculation. Lanaro and Patey\cite{Lanaro} also conducted large scale MD simulation with the SPC/E and JC force fields at similar supersaturation, and based on their reported induction time, their nucleation rate is consistent with the present result. 

There are few experimental studies of homogeneous nucleation of NaCl crystals from solution,\cite{Na, Gao, Desarnaud} and Zimmermann {\it et al.}\cite{Zimmermann} summarized these experimental estimates of nucleation rates in the Supplementary Information of Ref. \citenum{Zimmermann}. The SPC/E and JC force fields underestimate the nucleation rate by around 10 orders of magnitude compared to the experimental data from Gao {\it et al.}\cite{Gao} and Na {\it et al.}\cite{Na} Large discrepancies between simulation-based nucleation rates and experimental data have been observed for many nucleation processes, from crystallization of hard spheres\cite{Filion} to ice nucleation.\cite{Akbari1} The nucleation rate is sensitive to several thermodynamic quantities, especially the crystal/solution interfacial tension. Assuming the validity of classical nucleation theory, nucleation rate ($J$) is given by,
 
\begin{equation}
J = A e^{-\Delta G/k_b T}
\end{equation}
where $A$ is a kinetic prefactor and $\Delta G$, the nucleation free energy barrier, is given by
\begin{equation}
\Delta G = \frac{16 \pi \gamma^{3}}{3 \rho_s^2 \Delta \mu^2}
\end{equation}
$\gamma$ is the crystal/solution interfacial tension, $\rho_s$ is the number density of the crystal. $\Delta \mu$ is the chemical potential difference between the electrolyte and the crystal. In experimental estimation of nucleation rates,\cite{Desarnaud} the chemical potential difference is usually related to the mean ionic activity coefficient ($\gamma_{\pm}$), which may be extrapolated from the Pitzer's equation for supersaturated solution.\cite{Pitzer} At $S$ = 2.3, $\Delta \mu$ was estimated to be around 9.7 kJ/mol. For our simulations, following the method proposed by Mester and Panagiotopoulos,\cite{Mester} we estimated the chemical potential difference {\cor at} $S$ = 2.3 as 10.0 kJ/mol, in good agreement with the experimental estimate. The experimental NaCl crystal density is 2160 kg/m$^3$, while the calculated density from the JC force field is 2010 kg/m$^3$. It is difficult to measure directly the crystal/solution interfacial tension, and the value was experimentally estimated to be around 87 mN/m.\cite{Na} The prediction of interfacial tension from the JC and SPC/E force fields is not known and cannot be easily determined for a supersaturated solution. If we assume simulation data and experiment share the same kinetic prefactor ($A$) and the force fields overestimate the interfacial tension by 20\%, a deviation that is not unreasonable for empirical force fields, the nucleation rate would be underestimated by 9 to 10 orders of magnitude, which is similar to the deviation shown in Fig. \ref{rate}. While it is possible to explain the deviation between our simulation results and the experimental data from Gao {\it et al.}\cite{Gao} and Na {\it et al.}\cite{Na}, the probably more than 20 orders of magnitude underestimation of rates compared to the experimental data from Desarnaud {\it et al.} \cite{Desarnaud} may not be attributed to the overestimation of interfacial tension from the force fields. It is noted that the experimental measurements may be subject to significant systematic error, especially at lower supersaturations. For example, the presence of net charge in levitated droplets for nucleation measurement may promote crystallization\cite{Draper} and leads to overestimated nucleation rate.  

As mentioned in the Introduction, the difference {\cor in} chemical potential between crystal and equilibrium solution was underestimated in the work of Zimmermann {\it et al.} (blue triangles in the Fig. \ref{rate}),\cite{Zimmermann} as the correct equilibrium solubility of JC NaCl in SPC/E water is 3.7 mol/kg instead of 5.1 mol/kg. This in turn leads to an underestimation of interfacial tension and subsequent overestimation of nucleation rate. The critical nucleus size reported by Zimmermann {\it et al.}\cite{Zimmermann} are also much smaller than our results, with only part of the difference explainable by the use of different criteria for identifying crystalline structures. For example, at $S$ of 3.2, the size of the critical nucleus was 38 based on our FFS simulation while the seeding simulation from Zimmermann {\it et al.}\cite{Zimmermann} suggested such value was 6. In a recent simulation study, Lanaro and Patey\cite{Lanaro} found a nucleus with 10 ions has extremely low probability to survive for more than 6 ns in aqueous solution at $S$ = 4.2, which also questions the size of critical nucleus reported by Zimmermann {\it et al.}.\cite{Zimmermann} Overall, the origin of the differences between our results and those of Ref. \citenum{Zimmermann} is not entirely clear.

\subsection{Nucleation Pathway and Free Energy}
\begin{figure}[H]
\centering
\includegraphics[width=10cm]{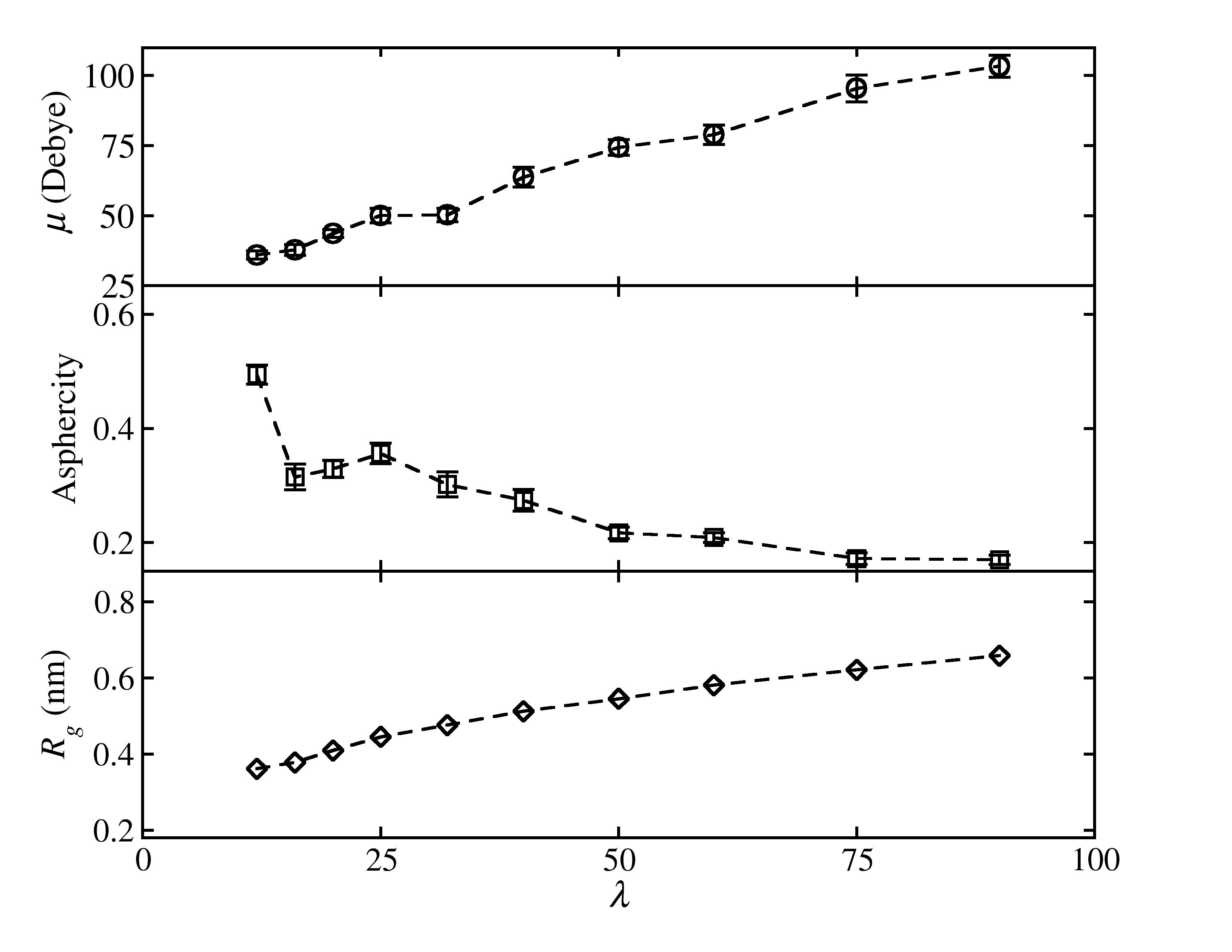}
\caption{Dipole moment ($\mu$, top), asphericity (middle) and radius of gyration ($R_g$, bottom) of crystalline nuclei of different size collected from FFS simulations at $S$=2.7. {\cor The dipole moment, asphericity, and radius of gyration were averaged over around 150 configurations collected at each milestone. The statistical uncertainty for the radius of gyration is smaller than the symbol size.}}
\label{prop}
\end{figure}

A large number of MD configurations were collected at each milestone of our FFS simulations (except the last two). Properties of crystalline nuclei and the nucleation pathway {\cor were} determined from such trajectories. Fig. \ref{prop} shows the dipole moment, asphericity ($\kappa$) and radius of gyration ($R_g$) of crystalline nuclei collected at different milestones ($\lambda$). As shown in the Fig. \ref{prop}, a crystalline nucleus has a non-zero dipole moment that increases with its size. While the perfect NaCl rock-salt (FCC) crystal has negligible dipole moment, a crystalline nucleus generally does not possess a perfect FCC structure thus has a non-zero dipole moment.  While it is expected that as the crystalline nucleus {\cor grows} further, its dipole moment should decrease as the internal part of the crystalline nucleus {\cor reconfigures} itself to a nearly perfect FCC structure, we did not observe a decrease of dipole moment, possibly because our FFS simulations mostly focused on early stages of nucleation and we did not collect many configurations for large crystalline nuclei that grew far beyond the critical size. The asphericity ($\kappa$) is defined as,
\begin{equation}
\kappa^2 = \frac{3}{2} \frac{\alpha_1^4+\alpha_2^4+\alpha_3^4}{(\alpha_1^2+\alpha_2^2+\alpha_3^2)^2} - \frac{1}{2}
\end{equation}
where $\alpha$ are the three real eigenvalues of the radius of gyration tensor. {\cor The value of asphericity varies between 0 (a spherical nucleus) and 1 (a linear nucleus).} As shown in Fig. \ref{prop}, a nucleus become more spherical {\cor as their size increases, and as expected the radius of gyration increases as the size of a nucleus increases.}

Since the observed dipole moments are quite high for larger nuclei, one concern is that a nucleus could interact with itself through the periodic boundaries, which in turn may affect the rates reported in the Fig. \ref{rate}. In order to check for possible finite-size effects on our rate calculations, we performed three FFS calculations of nucleation rates at $S$ = 3.2 using 500, 1000 and 3000 ions pairs, respectively. The nucleation rates obtained from these 3 simulations with different size agree with each other within statistical uncertainty, which {\cor suggests that} the finite size effect is not pronounced (data shown in Supplementary Material). The explanation for the lack of such effects despite the strong dipoles is that there is strong screening of the dipoles, both near the surface and at further distances from ions in solution at the high ionic concentrations that are of interest in this work. 

\begin{figure}[H]
\centering
\includegraphics[width=8cm]{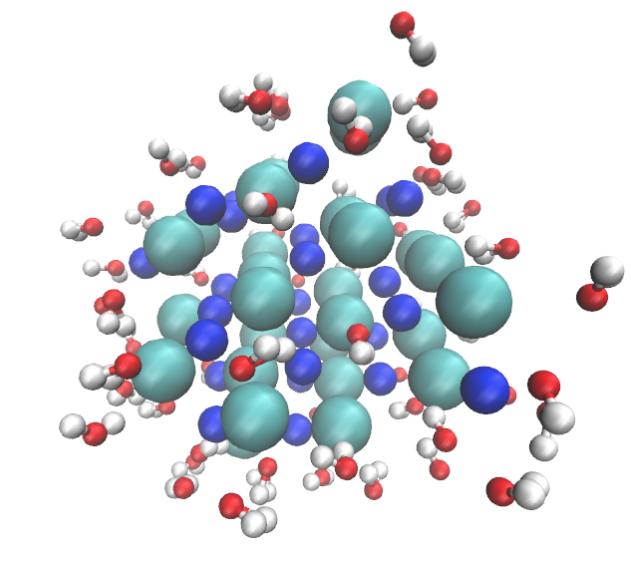}
\caption{A crystalline nucleus with 50 ions and its first hydration shell. Blue, cyan, red and white particles are Na$^+$, Cl$^-$, oxygen and hydrogen atoms, respectively.}
\label{snapshot}
\end{figure}

Fig. \ref{snapshot} shows a snapshot for a crystalline nucleus with 50 ions collected from FFS simulations at $S$=2.7. Water molecules that are within 0.35 nm of any ion in the nucleus were considered to be in the first hydration shell of the nucleus and are shown in Fig. \ref{snapshot}. It is clear that the nucleus has an ordered structure, and water molecules surround the nucleus, but do not penetrate its interior. Nuclei with different sizes and collected at different supersaturations show similar features. In order to quantify the structure of the crystalline nucleus, we used the octahedral ($\theta_{oct}$) and tetrahedral ($\theta_{tet}$) order parameter proposed by Zimmermann {\it et al.}\cite{Zimmermann} to investigate the local structure of ions (see Eqs. 3 and 4 in Supplementary Information of Ref. \citenum{Zimmermann}). The $\theta_{oct}$ ($\theta_{tet}$) compares the similarity between the local environment of an ion and the perfect rock salt/FCC (wurtzite/tetrahedral) structure. An ion was considered to have a rock salt/FCC (wurtzite/tetrahedral) structure if its $\theta_{oct}$ ($\theta_{tet}$) is larger than 0.15 while its $\theta_{tet}$ ($\theta_{oct}$) is smaller than 0.15. Using such criteria, it was found that all the ions in crystalline nucleus have FCC-like structure, except for the ones at the solid/solution interface. The same is true for the smaller pre-critical nuclei comprised of fewer than 20 ions. The presence of FCC-like structure in the crystalline nuclei, especially in nuclei at early stages of {\cor the nucleation process}, indicates the nucleation of NaCl in aqueous solution is consistent with the classical nucleation theory rather than the Ostwald step rule\cite{Santen}. As mentioned in the Introduction, Chakraborty and Patey\cite{Patey1, Patey2} observed {\cor that} nucleation of the NaCl crystal in solution follows the Ostwald step rule with aggregates of ions carrying large amount of water being intermediate between a uniform aqueous solution and the crystal. However, the supersaturations used in Refs. \citenum{Patey1} and \citenum{Patey2} were above 200, as the solubility of NaCl in water predicted from the OPLS force field was less than 0.02 mol/kg (details given in Supplementary Material). 

In addition to the size, the crystallinity of ionic clusters plays an important role in the nucleation process. We define crystallinity of a nucleus as the $\theta_{oct}$ parameter averaged over all its constituent ions. For  $S$=2.7, for instance, at the milestones of $\lambda$=16 and 25, we picked the 5 most crystalline and the 5 least crystalline configurations. From these 10 configurations, we initiated a large number of MD simulations by randomizing {\cor the velocities using} the Boltzmann distribution. As shown in Table \ref{table}, nuclei with high crystallinities have a higher transition probability {\cor of growing} (to the next milestone), compared to the transition probability estimated based on all the configurations collected at the milestone. It also took significantly longer for high-crystallinity nuclei to return back to the solution basin.
\begin{table}[H]
\begin{center}
\caption{Transition probability ($P$) and average time needed for a nucleus to return back to solution basin $t_{fail}$.}
\begin{tabular}{ccccc}
\hline
& \multicolumn{2}{c}{$\lambda$ = 16}   & \multicolumn{2}{c}{$\lambda$ = 25}  \\

                            & $P$  & $t_{fail}$ (ps) &  $P$  & $t_{fail}$ (ps)\\
\hline
5 high crystallinity confs  & 0.11   & $196\pm6$   & 0.09  &  $1200\pm12$   \\
5 low  crystallinity confs  & 0    & $48\pm3$    & 0     &  $515\pm15$   \\
all confs at the milestone  & 0.025  & $155\pm2$   & 0.02  &  $1000\pm13$ \\
\hline
\end{tabular}
\label{table}
\end{center}
\end{table}

\begin{figure}[H]
\centering
\includegraphics[width=10cm]{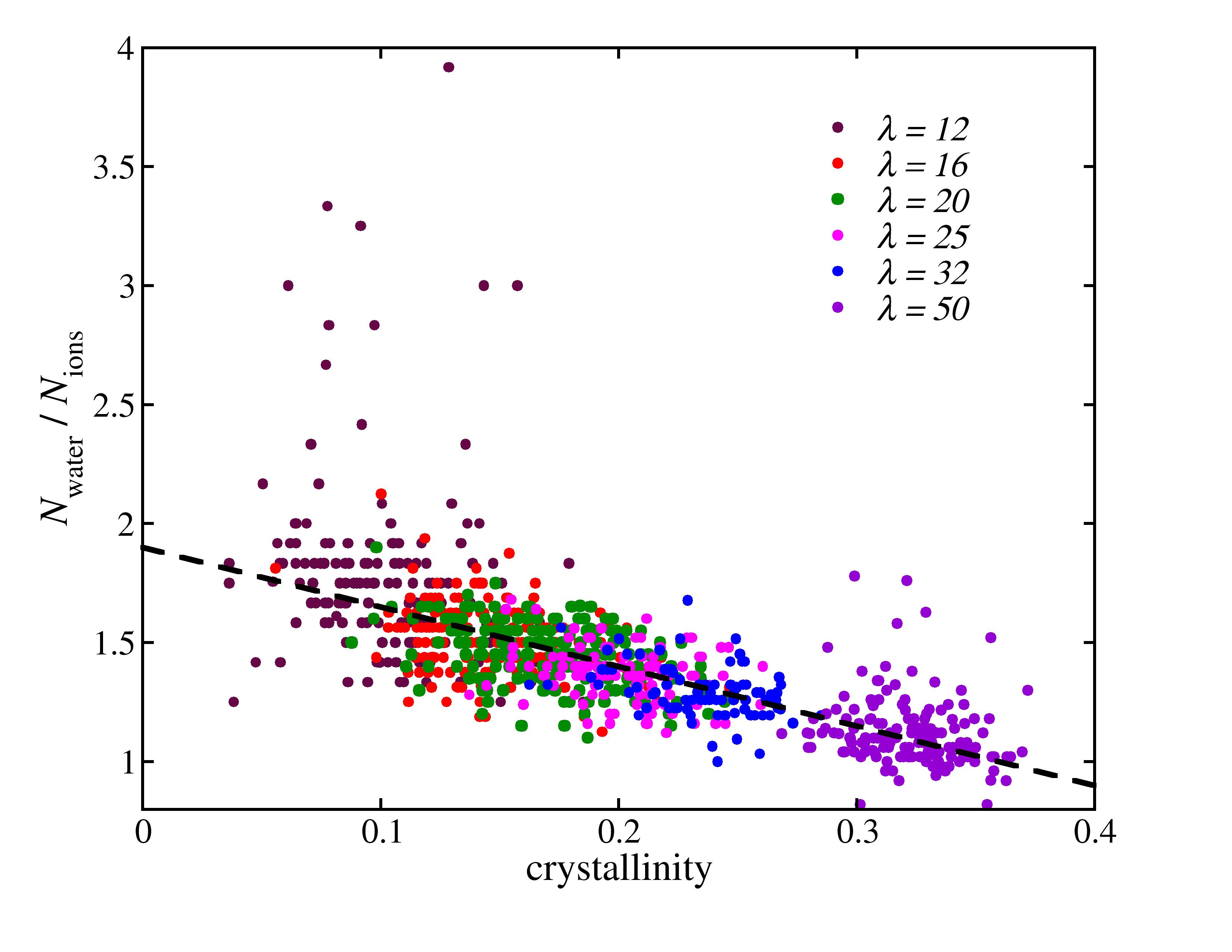}
\caption{Number of water molecules in the first hydration shell of a nucleus (nomarlized by the number of ions in nucleus) as a function of nucleus crystallinity. Each symbol corresponds to a crystalline nucleus collected at milestones of FFS simulations. The dashed line is a linear fit to the data.}
\label{water}
\end{figure}

Recently, Lanaro and Patey\cite{Lanaro} hypothesized that a nucleus with higher crystallinity may be less exposed to water. In Fig. \ref{water}, we show the number of water molecules in the first hydration shell of a crystalline nucleus (normalized by the number of ions in the nucleus) as a function of crystallinity. As shown in Fig. \ref{water}, large nuclei tend to have high crystallinity, as they {\cor becoming} more FCC-like as nucleation proceeds. Also, the amount of water near a nucleus decreases as the nucleus grows, shown by the negative slope of the linear fit. Similarly, for different nuclei with the same size (symbols with the same color), the amount of water exposed to a nucleus also decreases with crystallinity. This confirms the hypothesis of Lanaro and Patey\cite{Lanaro}, and may explain why higher-crystallinity nuclei are more likely to grow and have longer lifetime: such nuclei are less perturbed by surrounding water molecules.

\begin{figure}[H]
\centering
\includegraphics[width=10cm]{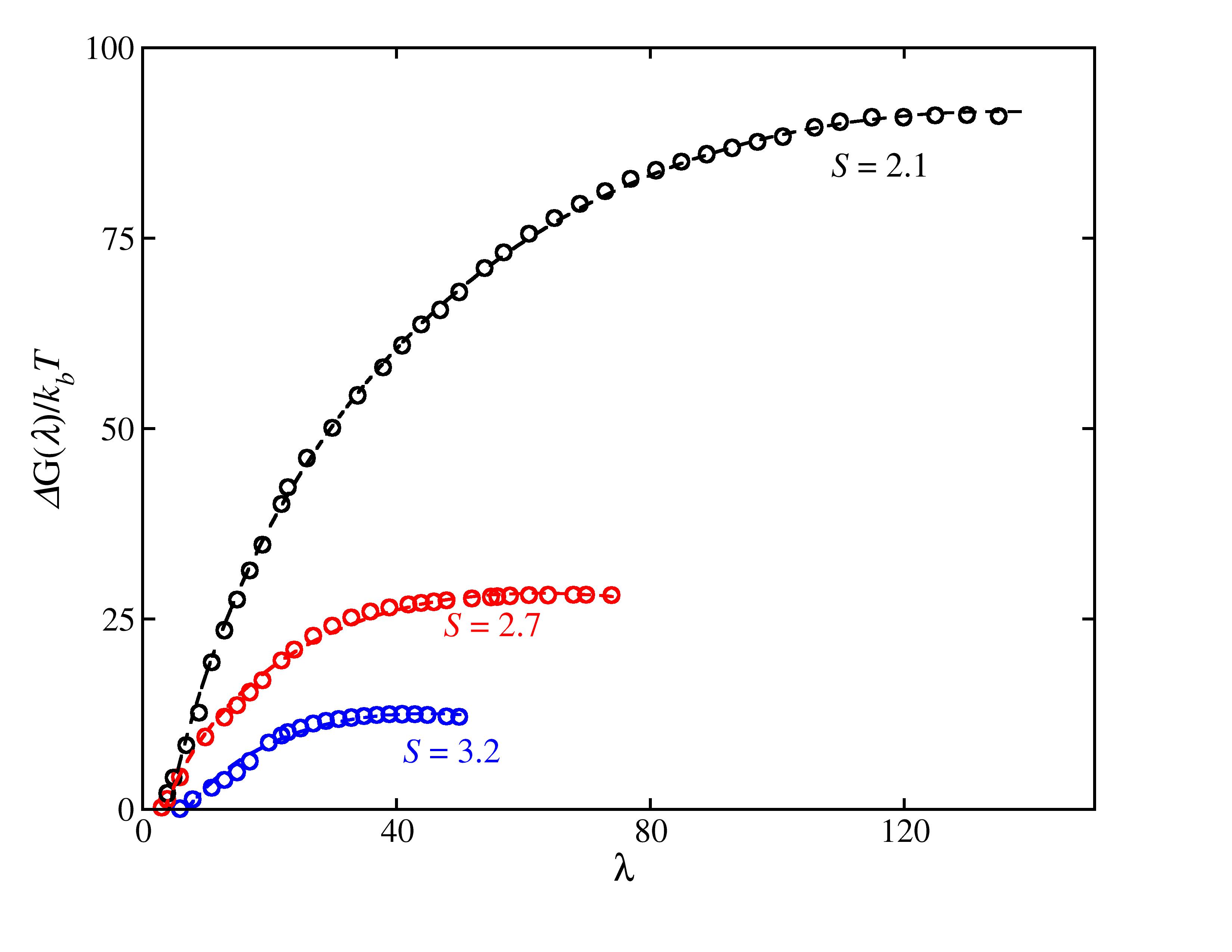}
\caption{Nucleation free energy profile at $S$ = 2.1, 2.7 and 3.2 {\cor estimated from the mean first passage time.\cite{Thapar}} Dashed lines are fitting to free energy profiles using Eqs. 4 and 5.}
\label{free}
\end{figure}

The nucleation free energy profile can be accessed from the mean first passage time (MFPT) and steady-state probability collected during FFS simulations using the algorithm recently proposed by Thapar and Escobedo.\cite{Thapar} As shown in the Fig. \ref{free}, the free energy barriers at $S$ = 2.1, 2.7 and 3.2 are around 91.0 $k_b T$, 28.0 $k_b T$ and 12.4 $k_b T$, respectively. Apparently, the free energy profiles show a single barrier at each supersaturation, indicating that the nucleation of NaCl in solution may not follow the two-step mechanism, in which a system {\cor must cross two free energy} barriers to reach crystallization. {\cor In order to compute additional quantities from the free energy barriers}, it was considered that the chemical potential of a crystalline nucleus equals the chemical potential of the rock-salt NaCl crystal, which is a reasonable approximation, especially for larger nuclei. Since the chemical potential difference ($\Delta\mu$) between solution and rock-salt/FCC crystal has been determined (details given in the Supplementary Material), the crystal/solution interfacial tension ($\gamma$) can be obtained from the Eq. 2 within the framework of classical nucleation theory. These values are listed in the Table \ref{gamma}.  Alternatively, the free energy profile of a nucleation process may be described with the following expression as suggested by classical nucleation theory,

\begin{equation}
\Delta G(\lambda) = - a \lambda + b \lambda^{2/3} + c
\end{equation}
and the $a$, $b$ and $c$ parameters can be determined by fitting the free energy profiles obtained from FFS simulations. Such {\cor fits} are shown as dashed lines in Fig. \ref{free}. The parameter $a$ corresponds to the chemical potential difference between solution and crystalline nucleus ($\Delta\mu' = 2a$). Note that {\cor this} chemical potential difference may not be equivalent to the chemical potential difference between solution and rock-salt crystal as small pre-critical nuclei, while {\cor showing} signatures of rock-salt/FCC structure, generally do not posses perfect rock-salt/FCC crystal structure. The $b$ parameter can be related to the crystalline nucleus/solution interfacial tension ($\gamma'$) as

\begin{equation}
\gamma' = \frac{b}{(9\pi \rho^{-2})^{1/3}}
\end{equation}
where $\rho$ is the density of a crystalline nucleus, which is assumed to be the density of rock-salt crystal from the JC force field (2010 kg/m$^3$). {\cor It is worth mentioning that}, unlike the crystal/solution interfacial tensions ($\gamma$) determined from Eq. 2 using the height of free energy barriers, the crystalline nucleus/solution interfacial tensions ($\gamma'$) were estimated using the entire free energy profile based on the Eq. 5, {\cor and could therefore be} interpreted as solid/solution interfacial tensions averaged over the entire nucleation pathway. The values of the crystal/solution ($\gamma$) and the path-averaged crystalline nucleus/solution ($\gamma'$) interfacial tensions are listed in the Table \ref{gamma}.  While experimental values of crystal/solution interfacial tension are not {\cor precisely} known, such value {\cor was estimated to be 87 mN/m using classical nucleation theory.\cite{Na} As shown in Table \ref{gamma}, the crystal/solution interfacial tensions, as well as the path-averaged crystalline nucleus/solution interfacial tensions, decrease as the supersaturation increases. The estimated crystal/solution interfacial tension is higher than the experimental estimate (87 mN/m), consistent with the underestimation of nucleation rates (see Figure \ref{rate}).} Except at $S$=2.1, the path-averaged interfacial tensions ($\gamma'$) are smaller than the solution/crystal interfacial tensions ($\gamma$), which is expected since at the early stages of the nucleation process, the interfacial tension between a small crystalline nucleus and its surrounding solution should be small. Such argument can be understood by taking the size of a crystalline nucleus to the {\cor small-nucleus} limit: the interfacial tension between a crystalline nucleus with zero ions, which becomes a part of solution, and its surrounding solution should be 0. At $S$=2.1, the critical nucleus size is much larger (120 ions), and the crystalline nucleus/solution interfacial tension averaged over the nucleation pathway becomes similar to the crystal/solution interfacial tension, as small nuclei constitutes a small part of the nucleation pathway.

\begin{table}[H]
\begin{center}
\caption{Chemical potential differences between rock-salt crystal and solutions ($\Delta\mu$, {\cor see SI}), fitting parameters $a$ and $b$, estimated interfacial tensions ($\gamma$) using Eq. 2, path-averaged interfacial tensions ($\gamma'$). }
\begin{tabular}{cccccc}
\hline
$S$ & $\Delta \mu$ (kJ/mol) & $2a$ (kJ/mol) &  $b$ (kJ/mol) & $\gamma$ (mN/m)  & $\gamma'$ (mN/m) \\
\hline
3.2 & 15.3  & 6.3  & 16.7  &  94.5    &  68.5  \\
2.7  & 12.6 & 6.6  & 19.8  &  108.8  &  81.3  \\
2.1  & 9.1   & 8.3  &  33.4 &  130.0  &  137.5 \\
\hline
\end{tabular}
\label{gamma}
\end{center}
\end{table}

{\cor Finally, it is worth mentioning that most of the computational cost of FFS simulations is associated with the sampling in the last few milestones. Long simulation times, typically longer than 5 ns, are required for large nuclei to grow to the next milestone or return to the solution basin. The FFS simulations {\cor and MD simulations of spontaneous nucleation} that produced nucleation rates at $S$ of 2.1, 2.7, 3.2, 4.1 and 4.5 took a total amount of 341 $\mu$s of MD trajectories, which corresponds to about 3.5 million CPU hours on Intel Haswell processors.}

\section{Conclusions}

The homogeneous nucleation of NaCl in supersaturated solution {\cor was} studied using {\cor the} forward flux sampling (FFS) method in {\cor conjunction with} molecular dynamics simulations. The nucleation rates were obtained for a range of supersaturations from 2.1 to 4.5, and the rates calculated from FFS were found to be consistent with those estimated from {\cor the} induction times for spontaneous nucleation. While the crystalline nucleus has a non-zero dipole moment that increases with nucleus size, finite size effect {\cor were} not found to be significant {\cor for} the calculation of nucleation rates. The crystallinity of nucleus plays an important role in the nucleation process. Nuclei with high crystallinity have higher {\cor probabilities} to grow and longer lifetimes, which is possibly because they have less water in their {\cor hydration} shells. From the trajectories collected at FFS milestones, we found that the aqueous NaCl nucleates directly into FCC-like rock-salt crystal, a process that is consistent with classical nucleation theory rather than {\cor with} Ostwald's step rule although prior simulation studies {\cor have} indicated the possibility of Ostwald step rule for the crystallization of NaCl in solution. {\cor This} discrepancy between nucleation mechanisms can {\cor plausibly} be attributed to the different NaCl solubilities using different force fields, {\cor as the prior simulations suggesting Ostwald step rule were conducted at unrealistically high supersaturation ($>$ 200) due to the severe underestimation of solubility from the OPLS and SPC/E force fields.} The performance of a force field with respect to its {\cor ability to predict} equilibrium properties (e.g. solubility) may significantly impact simulations of the nucleation process. The SPC/E and JC force fields underestimate the nucleation rates, compared to the available experimental data possibly due to their overestimation of crystal/solution interfacial tension. 

{\cor While the SPC/E and JC force fields show reasonable prediction for a series of thermodynamic properties of aqueous NaCl solution, and thus are widely used for simulations of electrolyte solutions, their underestimation/overestimation of nucleation rates/interfacial tension indicate intermolecular interactions are not captured satisfactorily by the force fields. Further development of accurate force fields that include advanced physics, such as polarization, is highly desirable in order to represent simultaneously kinetic (e.g. nucleation rates) and thermodynamic (e.g. solubility, activity coefficient) properties of aqueous electrolytes. While it has been shown that water and ions models that explicitly include polarization yield improved predictions for thermodynamic properties of aqueous electrolytes,\cite{pola} the performance of polarizable force fields for nucleation properties remains unknown and will be tested in future work.}

\section{Supplementary Material}
See supplementary material for parameters of SPC/E water and JC NaCl force fields, choice of milestones and transition probability of FFS simulations, chemical potentials as well as NaCl solubility of OPLS and JC models in SPC/E water.

\section{Acknowledgments}
Financial support for this work was provided by the Office of Basic Energy Sciences, U.S. Department of Energy, under Award DE-SC0002128. Additional support was provided by the National Oceanic and Atmospheric Administration (Cooperative Institute for Climate Science Award AWD 1004131). Calculations were performed on the Terascale Infrastructure for Groundbreaking Research in Engineering and Science (TIGRESS) computing facility at Princeton University. {\cor This work used the Extreme Science and Engineering Discovery Environment (XSEDE), which is supported by National Science Foundation grant number TG-CHE170059.} H.J thanks David Richard for helpful discussion on the free energy profile calculation.

\footnotesize{

}

\end{document}